\documentclass[preprint,showpacs,preprintnumbers,amsmath,amssymb,superscriptaddress,bibnotes, prl]{revtex4}

\usepackage{graphicx}
\usepackage{dcolumn}
\usepackage{tabularx}
\usepackage{bm}

\begin{document}

\preprint{Preprint}

\title{Optical control of magnetization of micron-size domains in antiferromagnetic NiO single crystals}

\author{Takuya Higuchi}
\affiliation{Department of Applied Physics, The University of Tokyo and CREST-JST, 7-3-1 Hongo, Bunkyo-ku, Tokyo 113-8656, Japan}
\author{Natsuki Kanda}
\affiliation{Department of Applied Physics, The University of Tokyo and CREST-JST, 7-3-1 Hongo, Bunkyo-ku, Tokyo 113-8656, Japan}
\author{Hiroharu Tamaru}
\affiliation{Photon Science Center, The University of Tokyo, Hongo, Tokyo 113-8656, Japan}
\author{Makoto Kuwata-Gonokami}
\email[E-mail: ]{gonokami@phys.s.u-tokyo.ac.jp}
\affiliation{Department of Applied Physics, The University of Tokyo and CREST-JST, 7-3-1 Hongo, Bunkyo-ku, Tokyo 113-8656, Japan}
\affiliation{Photon Science Center, The University of Tokyo, Hongo, Tokyo 113-8656, Japan}
\affiliation{Department of Physics, The University of Tokyo, Hongo, Tokyo 113-0033, Japan}
\date{\today}

\begin{abstract}
We propose Raman-induced collinear difference-frequency generation (DFG) as a method to manipulate dynamical magnetization.
When a fundamental beam propagates along a threefold rotational axis,
this coherent second-order optical process is permitted by angular momentum conservation through the rotational analogue of the Umklapp process. 
As a demonstration, we experimentally obtained polarization properties of collinear magnetic DFG along a [111] axis of a single crystal of antiferromagnetic NiO with micro multidomain structure, which excellently agreed with the theoretical prediction.
\end{abstract}


\maketitle

Recent developments in pulsed laser techniques have enabled ultrafast optical manipulation of magnetization in solids \cite{Kimel2005, Kalashnikova2008, Nishitani2010, Satoh2010, Bigot2009, Beaurepaire1996, Crooker1996, Kimel2004, Duong2004, Hansteen2005}.
In particular, coherent control through Raman-induced nonlinear optical processes \cite{Kimel2005, Kalashnikova2008, Nishitani2010, Satoh2010} has attracted interests, because low- energy magnetic excitations are accessible without excess heating \cite{Bigot2009}.
Therefore, compared with the conventional processes involving optical absorption followed by thermal relaxation \cite{Beaurepaire1996, Crooker1996, Kimel2004, Duong2004}, these coherent processes have potential applications including quantum information processing and ultrafast switching techniques for spintronics.

In such coherent optical processes, conservation of angular momentum determines the polarization of the induced excitation, 
whereas conservation of energy and momentum determines its frequency and wave vector, respectively \cite{Shen1966, Bloembergen1980}.
The balance of angular momentum includes contributions from the electromagnetic field (${\bf J}^{\rm EM}$), excitations in solids (${\bf J}^{\rm ex}$), and the crystalline lattice (${\bf J}^{\rm c}$): $\Delta {\bf J}^{\rm EM} + \Delta {\bf J}^{\rm ex} + \Delta {\bf J}^{\rm c} =0 $ \cite{Bloembergen1980}.
When the $z$-axis is a continuous rotational axis, $J_z^{\rm EM}+J_z^{\rm ex}$ must be rigorously conserved because $J_z^{\rm c}$ is a conserved quantity under any rotation around the $z$-axis.
By contrast, along a threefold rotational axis of the crystal, $J_z^{\rm EM}+J_z^{\rm ex}$ is conserved within a change of $3\hbar$; this is the rotational analogue of the Umklapp process \cite{Simon1968, Bloembergen1980}. 
This discrete rotational symmetry in solids, and the resultant polarization selection rules in nonlinear optics, has been taken into account for decades when the second-harmonic generation  \cite{Simon1968, Bloembergen1980} and the parametric down-conversion \cite{Visser2002} are discussed. 
However, little attention has been paid to implications of the discrete rotational symmetry for the coherent control of elementary excitations in solids.

In this Letter, we discuss selection rules for optically induced magnetization from a viewpoint of conservation of angular momentum.
The inverse Faraday effect \cite{Ziel1965} and the magnetic difference-frequency generation (DFG) are used as specific examples.
Both are described by third-rank axial $c$-tensors $\chi^{(2)\rm MEE}$ that change sign under time reversal \cite{Pershan1963, Birss-book}. 
The superscripts label magnetic (M) or electronic (E) interactions.
Subsequently, we propose collinear magnetic DFG along a threefold rotational axis
as a scheme of optical control of magnetization, using the discrete rotational symmetry.
As a demonstration, polarization selection rules are calculated and experimentally obtained for a collinear DFG in NiO that involves two frequency components of an ultrashort laser pulse, showing excellent agreement.

\begin{figure}[t]
\begin{center}
\includegraphics[width=8.5cm]{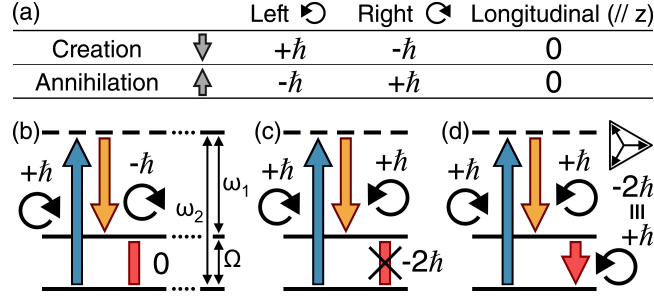}
\caption{\label{DFG-SFG} 
(Color online)
(a) Changes of $J^{\rm EM}_{z}$ by creation or annihilation of
photons propagating along the $z$-axis.
The lower panels show schematics of the changes of the angular momentum
in collinear scattering processes of photons
to a mode with (b) the same helicity and (c) the opposite,
and (d) a collinear DFG process allowed along a threefold axis involving two photons with opposite helicity.
}
\end{center}
\end{figure}

To discuss the conservation of angular momentum in light-matter interactions for photons propagating along the $z$-axis,
it is convenient to distinguish vector fields by their helicity.
Any electric field ${\bf E}$ with a propagating vector ${\bf k}$ along the $z$-axis can be decomposed into circular bases as:
${\bf E}(\omega, z) = 2{\rm Re}\left[\left( E_+ {\bf e}_+ +E_- {\bf e}_-  +E_z {\bf e}_z \right) e^{-i (\omega t-k z)}\right]$, with unit vectors, ${\bf e}_+=({\bf e}_x+i{\bf e}_y)/\sqrt{2}$, ${\bf e}_-=({\bf e}_x-i{\bf e}_y)/{\sqrt{2}}$, and ${\bf e}_z$.
$E_+$ and $E_-$ are left- and right- circular light fields, 
whereas $E_z$ is longitudinal.
As illustrated in Fig.~\ref{DFG-SFG}(a),
creation or annihilation of a photon induces $\Delta J_z^{\rm EM}=\pm \hbar$
in correspondence with its helicity, while $J_z^{\rm EM}=0$ for the longitudinal field along the $z$-axis.
The same rule is available for a magnetic field represented in circular bases.
Therefore, if and only if the angular momentum of an elementary excitation is $\pm \hbar$,
its transition to and from one photon that propagates along the $z$-axis is allowed.

A light beam at the frequency $\omega$ can induce dc magnetization ${\bf M}(0)$ through the inverse Faraday effect:
\begin{equation}
M_i^{(2)}(0)=\chi^{(2)\rm MEE}_{ijk}(0;-\omega,\omega) E_j^*(\omega)E_k(\omega),
\label{eq-IFE} \end{equation}
where ${\bf E}(\omega)$ is the electric field of the light wave \cite{Ziel1965}.
In a nonabsorbing isotropic medium invariant under time-reversal, $\chi^{(2)}_{ijk}=\chi \varepsilon _{ijk}$.
Here, $\chi$ is a scalar and $\varepsilon _{ijk}$ is the Levi-Civita tensor.
Then, Eq.~\eqref{eq-IFE} is reduced to ${\bf M}^{(2)}= \chi {\bf E}^* \times {\bf E} = \chi (|E_{+}|^2-|E_{-}|^2) {\bf e}_z$.
Thus, in an isotropic medium, an elliptically polarized light beam induces ${\bf M}^{(2)}$ along its propagating direction.

Similarly, ac magnetization ${\bf M}^{(2)}(\Omega)$ 
at the difference frequency $\Omega = \omega_1 - \omega_2$ can be parametrically generated 
by two light waves with frequencies $\omega_1$ and $\omega_2$ through a stimulated Raman-type nonlinear process \cite{Yan1985, Shen-book}:
\begin{equation}
M^{(2)}_{i}(\Omega) = \chi^{(2) \rm MEE}_{ijk} (\Omega; -\omega_1, \omega_2)
E_{j}^*(\omega_1) E_{k}(\omega_2). \label{eq-magnetic material excitation}
\end{equation}
When ${\bf M}^{(2)}(\Omega)$ is not only Raman (two-photon) active but also infrared (one-photon) active (via a magnetic dipole process), 
magnetic field ${\bf H}^{(2)}(\Omega)$ and electric field ${\bf E}^{(2)}(\Omega)$ are radiated
from ${\bf M}^{(2)}(\Omega)$ \cite{Shen-book}:
\begin{eqnarray}
\left[ \nabla \times ( \nabla \times ) - \frac{\Omega^2}{c^2}\varepsilon_{\rm r} \right]{\bf H}^{(2)}(\Omega) 
&=&
\frac{\Omega^2}{c^2}\varepsilon_{\rm r} {\bf M}^{(2)}(\Omega) 
\label{eq-HM} \\
- i \Omega  \varepsilon _r \varepsilon _0  {\bf E}^{(2)}(\Omega)
&=& 
\nabla \times {\bf H}^{(2)}(\Omega), \label{eq-HE}
\end{eqnarray}
where $\varepsilon_{\rm r}$ is the relative permittivity at $\Omega$, 
$\varepsilon_0$ is the vacuum permittivity,
and $c$ is the speed of light; this is the magnetic DFG.

In continuous rotational symmetry, as in a liquid or gas,
the requirement of rigorous conservation of $J_z^{\rm EM}+J_z^{\rm ex}$ forbids collinear DFG --- i.e., DFG with incident and out-going photons propagating along the same ($z$-) direction.
In particular, optically induced magnetic excitation in the medium has $J_z^{\rm ex}$ of either 0 or $\pm 2 \hbar$, and is thus nonradiative [Figs. \ref{DFG-SFG}(b) and (c)].
Note that generation of longitudinal field ($J_z^{\rm ex}=0$), rather than DFG, is possible
by scattering the incident photon to a mode with the same helicity.
If the scattered mode is the identical one with the same frequency,
the result is a dc field along the $z$-axis, which contributes to the inverse Faraday effect. 

However, DFG is not necessarily forbidden if a fundamental light beam 
propagates along a crystal's threefold symmetry axis
and involves two photons of opposite helicity --- a linearly polarized pulse satisfies this condition, while purely circularly polarized pulse does {\it not}.
This is because this discrete symmetry allows the change of $J_z^{\rm c}$ by $3 \hbar$ \cite{Bloembergen1980, Visser2002}. 
For example, annihilation of the right-circular photon and creation of the left-circular photon
induces a magnetic excitation with $J_z^{\rm ex}=-2\hbar$. 
This is equivalent to $+\hbar$ through decrease of $J_z^{\rm c}$ by $3 \hbar$,
and emits a left-circular photon [Fig.~\ref{DFG-SFG}(d)].
The result is that $J_z^{\rm EM}$ increases by $3 \hbar$ and the total angular momentum is conserved.
For an incident pulse that is linearly polarized, a mirror process occurs simultaneously, where $\Delta J_z^{\rm EM} = -3 \hbar$ and $\Delta J_z^{\rm c} = +3 \hbar$.

Nonvanishing transverse components of $\chi^{(2) \rm MEE}$ invariant under rotation by $2\pi/3$ around $z$-axis
describes the allowed collinear DFG process along a threefold axis:
\begin{eqnarray}
\chi_{xxx}&=&-\chi_{xyy}=-\chi_{yxy}=-\chi_{yyx} \equiv \alpha , \label{eq-alpha}\\
\chi_{yyy}&=&-\chi_{yxx}=-\chi_{xyx}=-\chi_{xxy} \equiv \beta, \label{eq-beta} 
\end{eqnarray}
and other components vanish.
Here, $\alpha$ and $\beta$ may also be zero depending on additional spatial or temporal symmetry operations allowed.
In the circular basis, they are
\begin{eqnarray}
\chi_{+++} &\equiv& \chi_{ijk}({\bf e}_+)^*_i({\bf e}_+)^*_j({\bf e}_+)^*_k = \sqrt{2}(\alpha + i \beta) \\ \chi_{---}&\equiv&  \chi_{ijk}({\bf e}_-)^*_i({\bf e}_-)^*_j({\bf e}_-)^*_k = \sqrt{2}(\alpha - i \beta),
\end{eqnarray}
which correspond to the processes that involve change of $J_z^{\rm EM}$ by $3\hbar$ and $-3\hbar$, respectively.
When the amplitudes of these two processes are the same, as in the case of NiO discussed below, the resultant radiation is linearly polarized for an incident pulse with any fixed polarization.

In the remaining paragraphs, the above description of the collinear DFG 
will be applied to the interpretation of our experimental observation of THz radiation from  antiferromagnetic magnons in NiO illuminated by laser pulses.
We chose the direction of the propagation of light 
to be parallel to the [111] axis of an as-grown crystal of NiO that consists of randomly distributed domains due to magnetic ordering.
This axis is effectively threefold if the signal induced in each domain is coherently superposed, as we will discuss below.
NiO has a magnon resonance around 1 THz that is both Raman \cite{Lockwood1992} and  infrared \cite{Kondoh1960, Sievers1963} active.
Recently, researchers reported radiation from the magnon \cite{Nishitani2010} and oscillation of the Faraday rotation  \cite{Satoh2010} after illumination by pulsed lasers.
Our setup differs from these previous studies in the selection of the crystal orientation and the domain structure because it is crucial for a clear demonstration of this DFG process.
Note also that the absence of spontaneous magnetization in NiO rules out photomagnetic effect \cite{Hansteen2005}.

\begin{figure}[t]
\begin{center}
\includegraphics[width=8.5cm]{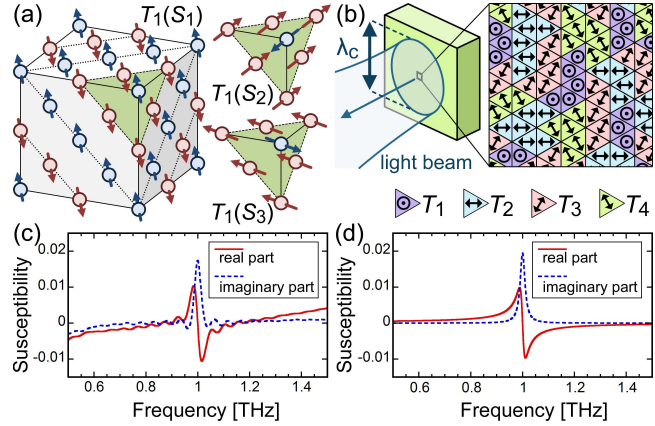}
\caption{\label{NiO-domain} 
(Color online)
(a) Schematics of spin alignment in three $S$-domains belonging to the $T_1$-domain,
and (b) random distribution of four orientations of $T$-domains within the coherent length $\lambda_{\rm C}$ of the light beam.
(c) Experimental and (d) calculated magnetic susceptibility of NiO;  $\Gamma = 70$ GHz and $\varepsilon_r = 12.25$ are assumed.}
\end{center}
\end{figure}

NiO crystallizes in a rocksalt structure above its N\'eel temperature of $T_{\rm N}=523$ K.
Below $T_{\rm N}$, the Ni$^{2+}$ ions with opposite sign of spins ($S=1$) occupy equivalent crystallographic positions [Fig.~\ref{NiO-domain}(a)], forming two sublattices.
Subsequent lattice distortion induces four variants called $T$-domains, and each of them have three variants called $S$-domains
that differ by their spin orientation (total of 12 variants) \cite{Slack1960}.
For one of the $T$-domains, $T_1$, the crystal shows slight rhombohedral distortion along its $[111]$ axis; for $T_2$ to $T_4$, the distortion is along the other three diagonals.
The domains are accommodated by twin structures randomly distributed in an as-grown NiO crystal [Fig.~\ref{NiO-domain}(b)] \cite{Sanger2006},
and their sizes were several micrometers according to polarization microscopy. 
The spin orientation in one of the three $S$-domains belonging to $T_1$-domains (denoted by $T_1(S_1)$) is $[11\bar2]$
and for $T_1(S_2)$ and $T_1(S_3)$, their equivalents in the $(111)$ plane [Fig.~\ref{NiO-domain}(a)].

To examine whether all domains coherently contribute to the magnetic response, we measured the
magnetic susceptibility tensor of an as-grown NiO crystal near its magnon resonance ($\sim$ 1 THz) by time-domain spectroscopy at room temperature  [Fig.~\ref{NiO-domain}(c)].
Ellipsometric measurements showed that the magnetic response is isotropic \cite{Kanda2007}.
This indicates that although the susceptibility of each $S$-domain is anisotropic \cite{Nagamiya1951, Kittel1951}, 
coherent superposition of the 12 variants results in effectively isotropic susceptibility, described by a scalar $\chi_{\mu}(\omega)$.

The magnetic susceptibility of this magnon resonance is calculated as: 
$ \chi_{\mu}(\omega)= 
\frac{2}{3}
\frac{2\gamma ^2 M H_{\rm A}}{\gamma ^2 H_{\rm A}(2 A M + H_{\rm A})-\omega^2 + 2 i \Gamma \omega} \label{eq-susceptibility}$,
where $\gamma$ is the gyromagnetic ratio of an electron, $\Gamma$ is the damping coefficient, $M$ is the sublattice magnetization, and
$A M$ and $H_{\rm A}$ are the exchange and anisotropy fields, respectively \cite{Nagamiya1951, Kittel1951}. 
The factor of $2/3$ is the result of averaging over the domains.
Figure \ref{NiO-domain}(d) is the plot of $ \chi_{\mu}(\omega)$ using 
$M$, $A$, and $H_{\rm A}$ from the literature \cite{Hutchings1972} and
taking thermal suppression of $M$ into account \cite{Kondoh1960, Sievers1963}.
The quantitative agreement with the experimental result assured us that the $S$-domains are sufficiently large to accommodate magnons.

\begin{figure}[t]
\begin{center}
\includegraphics[width=8.5cm]{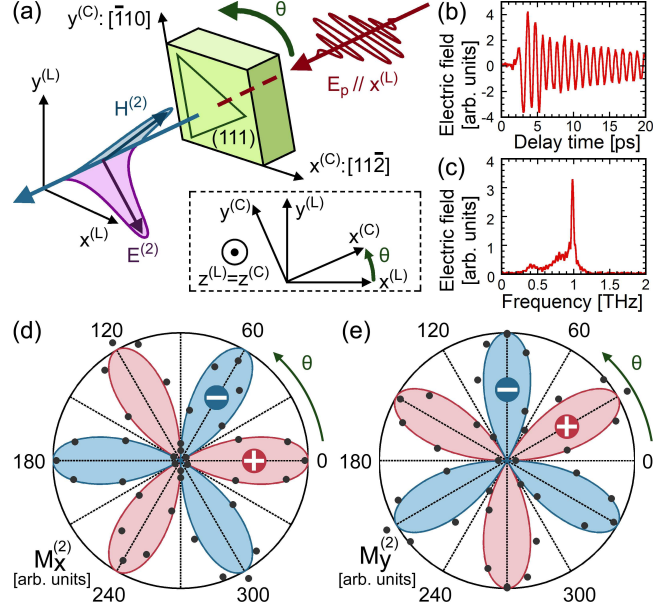}
\caption{\label{polar-merged} 
(Color online)
(a) Schematics of the experimental setup. Inset shows the relation between crystal and laboratory coordinates.
(b) Temporal waveform of the radiation from NiO and (c) its Fourier transformed spectrum.
Polar plots of the transverse components (d) $M_{x}^{(2)}$ and (e) $M_{y}^{(2)}$ of the induced magnetization as functions of the crystal rotation angle $\theta$.
The dots are experimental data of their amplitude,
and the solid curves are calculated by Eq.~\eqref{eq-polarization NiO}.
}
\end{center}
\end{figure}

The DFG in this sample, therefore, is also described by coherent superposition of the signal generated in each of the 12 variants:
\begin{equation}
M^{(2)}_i  = \chi^{(2)\rm MEE}_{ijk} E_j^* E_k \equiv \sum _n \frac{1}{12} \chi^{(2){\rm MEE},n}_{ijk} E_j^* E_k .
\end{equation}
Here, $\chi^{(2){\rm MEE},n}$ is the nonlinear susceptibility tensor of an $n$-th domain,
which can be characterized by its orientation with respect to the crystal's Cartesian frame, denoted by superscript (C), as depicted in Fig.~\ref{polar-merged}(a).
In a single $T_1(S_1)$-domain, 
the allowed symmetry operations are
$ 1$, $\overline{1}$, $\underline{2_y}$, and $\underline{\overline {2_y}}$, and magnetic point group of a Ni site is $ \underline{2} / \underline{\rm m} $
(the underbars denote time-reversal operation) \cite{Birss-book}.
This gives 14 nonvanishing independent components of $\chi^{(2){\rm MEE},n}$,
including $\chi_{zzz}$, $\chi_{xxx}$, $\chi_{yyx}$, $\chi_{zzx}$, $\chi_{xxz}$, $\chi_{yyz}$,
and other components obtained by permutation of the Cartesian indices.
Tensors for other variants are obtained,
as is discussed for magnetic second-harmonic generation \cite{Sanger2006}:
the tensor of a $T_1(S_2)$-domain is calculated by transforming $\chi^{(2){\rm MEE},S_1}$ under rotation by $2\pi/3$ around the $z^{(\rm C)}$-axis.
The nonvanishing transverse components of $\chi^{(2)\rm MEE}$ in the crystal coordinate frame is derived by averaging $\chi^{(2){\rm MEE},n}$ over the 12 variants as:
\begin{equation}
\chi_{xxx} = -\chi_{xyy} = -\chi_{yxy} = -\chi_{yyx}=\alpha . \label{eq-crystal coordinate}
\end{equation}
These are precisely the transverse components of a tensor that is invariant under rotation by $2\pi/3$: 
a special case of Eqs. \eqref{eq-alpha} and \eqref{eq-beta} in which $\beta = 0$.
Therefore, this multidomain crystal behaves in the same way as a single-domain crystal with threefold symmetry as  far as the DFG is concerned,
and thus collinear DFG along the $z^{\rm (C)}$-axis is allowed.

We performed time-domain spectroscopy measurements of THz radiation from NiO.
The crystal was irradiated by Ti-sapphire-based ultrashort laser pulses ($\sim$ 100 fs) along its [111] crystallographic axis, as shown in Fig.~\ref{polar-merged}(a).  
With linearly polarized excitation, the DFG was observed, but not with circular polarization.
Figures \ref{polar-merged}(b) and (c) show
a typical temporal waveform of the generated THz field ${\bf E}^{(2)}(t)$, and its Fourier transformed spectrum $|{\bf E}^{(2)}(\Omega)|$, respectively.
Radiation with a resonant frequency of $\Omega_0 \sim 1$ THz was observed.
Note that the absorption coefficient of NiO is small (60 cm$^{-1}$) 
for the laser wavelength ($\sim800$ nm) \cite{Newman1959}, 
and thus, the theory for nonabsorbing media is applicable.

Polarization properties of the THz radiation were observed by THz ellipsometry.
Incident light ${\bf E}_{\rm p}$ was fixed to be $x^{\rm (L)}$-polarized, and the sample was rotated by an angle $\theta$ around the $z^{\rm (L)}$-axis:
the superscript (L) denotes the laboratory coordinate, as shown in Fig.~\ref{polar-merged}(a).
The radiation was linearly polarized for any $\theta$,
and the $x^{\rm (L)}$- and $y^{\rm (L)}$- polarized components of ${\bf E}^{(2)}(t)$ were acquired using a wire-grid polarizer installed behind the sample \cite{Kanda2007}.
Transverse components of ${\bf M}^{(2)}(\Omega_0)$ at the resonant frequency $\Omega_0$ were determined as functions of $\theta$, using Eqs.~\eqref{eq-HM} and \eqref{eq-HE}, and are shown in Figs.~\ref{polar-merged}(d, e).

In the laboratory coordinate frame, $M_x^{(2)}$ and $M_y^{(2)}$ can be calculated by Eq.~\eqref{eq-magnetic material excitation},
in which the susceptibility tensor is derived by rotating the tensor in Eq.~\eqref{eq-crystal coordinate} by $\theta$ around the $z^{\rm (L)}$-axis:
\begin{equation}
M_x^{(2)} = \alpha |{\bf E}_{\rm p}|^2 \cos 3\theta ,~ 
M_y^{(2)} = \alpha |{\bf E}_{\rm p}|^2 \sin 3\theta . \label{eq-polarization NiO}
\end{equation}
Here, ${\bf E}_{\rm p} \parallel {\bf e}_x^{\rm (L)}$ is used.
Angular dependence of Eq.~\eqref{eq-polarization NiO} is depicted in Figs.~\ref{polar-merged}(d, e),
showing excellent agreement with the experimental data.
In addition, the measured amplitude $|M^{(2)}|$ was proportional to $|{\bf E}_{\rm p}|^2$,
indicating that the radiation mechanism is indeed a second-order nonlinear optical process.
It is important that the THz ellipsometry along a linearly isotropic axis allows us to obtain directly the orientation of the induced magnetization in contrast with nonlinear optical techniques \cite{Woodford2007}; for example by the Faraday rotation \cite{Kimel2004, Kimel2005, Kalashnikova2008, Satoh2010}, the Kerr rotation \cite{Beaurepaire1996, Bigot2009}, or second-harmonic generation \cite{Duong2004}.

In summary, we showed that magnetization can be induced through 
a collinear magnetic DFG process, which is allowed with linearly polarized excitation along a threefold axis.
The rotational analogue of the Umklapp process plays a key role because it allows the change of $J_z^{\rm EM}+J_z^{\rm ex}$ by $\pm 3\hbar$.
We experimentally demonstrated the DFG resonant on the antiferromagnetic magnons in a multidomain NiO crystal. 
We observed clear polarization dependence on the incident and radiated photons under linearly polarized excitation,
in excellent agreement with the theoretical description.
It has highlighted that the conservation of angular momentum can provide a general guiding principle to design experiments by taking material's symmetries into account.
Note that this concept is readily applicable to other elementary excitations induced through the nonlinear optical processes, for example, electronic DFG or optical rectification because both electric and magnetic fields behave in the same way under 
a rotation that maintains the parity (proper rotation).
Extensive applications of this approach to other systems,
such as artificial structures or atomic gasses in optical lattices, are expected.

The authors appreciate Yu. P. Svirko for his critical reading of the manuscript,
Y. Tanabe, K. Yoshioka, K. Konishi, and T. Shimasaki for their fruitful discussions, 
and H. Shimizu for experimental support.
This work was supported by KAKENHI 20104002 and 21020011
and by the Photon Frontier Network Program of MEXT, Japan.


\begin{thebibliography}{100}

\bibitem{Kimel2005}
A.~V. Kimel {\it et~al.}, Nature {\bf 435},  655  (2005).

\bibitem{Kalashnikova2008}
A.~M. Kalashnikova {\it et~al.}, Phys. Rev. B {\bf 78},  104301  (2008).

\bibitem{Nishitani2010}
J. Nishitani {\it et~al.}, Appl. Phys. Lett. {\bf
  96},  221906  (2010).

\bibitem{Satoh2010}
T. Satoh {\it et~al.}, Phys. Rev. Lett. {\bf 105},  077402  (2010).

\bibitem{Bigot2009}
J.-Y. Bigot, M. Vomir, and E. Beaurepaire, Nat. Phys. {\bf 5},  515  (2009).

\bibitem{Beaurepaire1996}
E. Beaurepaire {\it et~al.}, Phys. Rev. Lett. {\bf
  76},  4250  (1996).

\bibitem{Crooker1996}
S.~A. Crooker {\it et~al.}, Phys. Rev. Lett. {\bf 77},  2814  (1996).

\bibitem{Kimel2004}
A.~V. Kimel {\it et~al.}, Nature {\bf 429},  850  (2004).

\bibitem{Duong2004}
N.~P. Duong, T. Satoh, and M. Fiebig, Phys. Rev. Lett. {\bf 93},  117402
  (2004).

\bibitem{Hansteen2005}
F. Hansteen {\it et~al.}, Phys. Rev. Lett. {\bf 95},
  047402  (2005).

\bibitem{Shen1966}
Y.~R. Shen and N. Bloembergen, Phys. Rev. {\bf 143},  372  (1966).

\bibitem{Bloembergen1980}
N. Bloembergen, J. Opt. Soc. Am. {\bf 70},  1429  (1980).

\bibitem{Simon1968}
H.~J. Simon and N. Bloembergen, Phys. Rev. {\bf 171},  1104  (1968).

\bibitem{Visser2002}
J. Visser, E.~R. Eliel, and G. Nienhuis, Phys. Rev. A {\bf 66},  033814
  (2002).

\bibitem{Ziel1965}
J.~P. van~der Ziel, P.~S. Pershan, and L.~D. Malmstrom, Phys. Rev. Lett. {\bf
  15},  190  (1965).

\bibitem{Pershan1963}
P.~S. Pershan, Phys. Rev. {\bf 130},  919  (1963).

\bibitem{Birss-book}
P.~R. Birss,  in {\em Symmetry and Magnetism, Selected Topics in Solid State
  Physics, Vol. III}, edited by E.~P. Wohlfarth (North-Holland, Amsterdam,
  1966).

\bibitem{Shen-book}
Y.~R. Shen, {\em The principles of nonlinear optics} (Wiley-Interscience,
  Hoboken, 2003).

\bibitem{Yan1985}
Y.-X. Yan {\it et~al.}, J. Chem. Phys. {\bf 83},  5391  (1985).

\bibitem{Lockwood1992}
D. Lockwood, M. Cottam, and J. Baskey, J. Magn. Mag. Mater. {\bf 104},  1053
  (1992).

\bibitem{Kondoh1960}
H. Kondoh, J. Phys. Soc. Jpn {\bf 15},  1970  (1960).

\bibitem{Sievers1963}
A.~J. Sievers and M. Tinkham, Phys. Rev. {\bf 129},  1566  (1963).

\bibitem{Slack1960}
G.~A. Slack, J. Appl. Phys. {\bf 31},  1571  (1960).

\bibitem{Sanger2006}
I. S\"anger {\it et~al.} Phys. Rev. B {\bf 74},
  144401  (2006).

\bibitem{Kanda2007}
N. Kanda, K. Konishi, and M. Kuwata-Gonokami, Opt. Exp. {\bf 15},  11117
  (2007).

\bibitem{Nagamiya1951}
T. Nagamiya, Prog. Theor. Phys {\bf 6},  342  (1951).

\bibitem{Kittel1951}
C. Kittel, Phys. Rev. {\bf 82},  565  (1951).

\bibitem{Hutchings1972}
M.~T. Hutchings and E.~J. Samuelsen, Phys. Rev. B {\bf 6},  3447  (1972).

\bibitem{Newman1959}
R. Newman and R.~M. Chrenko, Phys. Rev. {\bf 114},  1507  (1959).

\bibitem{Woodford2007}
S.~R. Woodford, A. Bringer, and S. Blugel, J. Appl. Phys. {\bf 101},  053912
  (2007).

\end{thebibliography}

\newpage

\end{document}